\documentstyle[epsf,aps]{revtex}
\begin{document}
\newcommand{\be}{\begin{eqnarray}}
\newcommand{\dlq}{\lq\lq}
\newcommand{\ee}{\end{eqnarray}}
\newcommand{\ben}{\begin{eqnarray*}}
\newcommand{\een}{\end{eqnarray*}}
\newcommand{\stackeven}[2]{{{}_{\displaystyle{#1}}\atop\displaystyle{#2}}}
\newcommand{\lsim}{\stackeven{<}{\sim}}
\newcommand{\gsim}{\stackeven{>}{\sim}}
\renewcommand{\baselinestretch}{1.0}
\newcommand{\as}{\alpha_s}
\def\eq#1{{Eq.~(\ref{#1})}}
\def\fig#1{{Fig.~\ref{#1}}}
\begin{flushright} 
NT@UW--01--018
\end{flushright}
\vspace*{1cm} 
\setcounter{footnote}{1}
\begin{center}
{\Large\bf Diffractive Gluon Production in Proton--Nucleus Collisions
\\[.5cm] and in DIS}
\\[1cm]
Yuri V. Kovchegov\\ ~~ \\
{\it Department of Physics, University of Washington, Box 351560}
\\ {\it Seattle, WA 98195-1560 }\\ ~~ \\ ~~ \\
\end{center}
\begin{abstract} 
We derive expressions for the gluon production cross sections in the
single diffractive proton--nucleus scattering and DIS processes in a
quasi-classical approximation. The resulting cross sections include
the effects of all multiple rescatterings in the classical background
field of the target proton or nucleus, which remains intact after the
scattering. We also write down an expression for the inclusive gluon
production cross section in DIS in the quasi-classical approximation.
\end{abstract}

\section{Introduction} 

In the recent years it has become clear that partonic saturation may
play an important and possibly vital role in high energy hadronic and
nuclear collisions. Partonic saturation, associated with slowing
down of the growth of structure functions with energy, is
characterized by high density of quarks and gluons and strong gluonic
fields in the hadronic and nuclear wave functions
\cite{glrmq,mv,yuri,jklw}. The transition to the saturation region is 
described by the saturation scale $Q_s^2 (s)$, which is an increasing
function of the center of mass energy $s$ and for high enough $s$
could lie in the perturbative QCD region \cite{mv,me,bal}. It has been
suggested that most of the quarks and gluons in the small-$x$ hadronic
wave function have the transverse momentum of the order of $Q_s$
\cite{musat}, making perturbative calculation possible for most
observables in high energy hadronic collisions.

Calculations in the saturation region consist of two stages: classical
and quantum. As it was conjectured in \cite{mv} the strong gluonic
fields in the small-$x$ tail of a hadronic or nuclear wave function
could be very well approximated by the classical solution of the
Yang-Mills equations of motion. Even though this conjecture was
originally formulated with the valence quarks as sources of color
charge generating the gluon field \cite{mv} it could be generalized to
include all the higher momentum components of the nuclear wave
function in the source current \cite{JKLW}. The small-$x$ gluons have
a large coherence length of the order of $1/2 m_N x$ allowing them,
for small enough $x$, to coherently interact with all the nucleons in
the nucleus \cite{fs}. The classical gluonic field of a nucleus, known
as the non-Abelian Weizs\"{a}cker-Williams field, was shown to include
all the multiple rescatterings of the gluons with the sources of color
charge \cite{yuri,jklw}. In the valence quark formulation of the
problem for the nucleus where the multiple rescatterings are mediated
by two gluon exchanges this parametrically corresponds to resumming
all powers of the parameter $\as^2 A^{1/3}$, with $A$ the atomic
number of the nucleus \cite{yuri}. In terms of dynamical variables
resummation of this parameter is equivalent to summing up powers of
$Q_s^2 / k_\perp^2$, where $k_\perp$ is the transverse momentum of the
gluon. The physical effect of multiple rescatterings is to push the
gluon's transverse momentum from some almost non-perturbative scale up
to $k_\perp \sim Q_s$ thus suppressing the infrared region and the
phenomena associated with it \cite{kkl2}.

The second stage of the saturation calculations corresponds to
inclusion of quantum corrections. Usually the quantum corrections are
taken in the leading logarithmic approximation and bring in powers of
$\as \ln 1/x$, though in principle subleading logarithmic
contributions should be considered, since while being parametrically
smaller they could still be of numerical importance \cite{NLO}. In the
traditional language the leading logarithmic corrections correspond to
resummation of multiple BFKL pomeron exchanges \cite{BFKL,glrmq}. They
can also be represented as a series of emissions of the non-Abelian
Weizs\"{a}cker-Williams field, where after each interaction the
produced field is incorporated in the hard source for production of
softer field at the next step \cite{JKLW}.

Several observables have been calculated using the saturation
approach. The one which probably received most attention is the total
cross section of deep inelastic scattering (DIS), since it is related
to quark and gluon distribution functions. In the ``classical''
approximation it has been first calculated for QCD in \cite{Mue},
yielding a Glauber-type formula. Then an extensive effort was put into
inclusion of the quantum corrections in the total DIS cross section
and the gluon distribution function resulting in the nonlinear
equation of \cite{me,bal}. The equation could be derived in high
energy effective theory of \cite{bal} or using the dipole model of
\cite{dip}. Later the nonlinear equation has been reproduced and
solved by approximate and numerical methods in \cite{rep}.

Another important observable which could be calculated in the
saturation framework is the inclusive single gluon production cross
section in DIS, proton-nucleus (pA) and nucleus-nucleus (AA)
collisions. The gluon production cross section could be compared with
the data on minijet and pion production at mid-rapidity. In the first
stage of the saturation calculation the problem of finding the
inclusive single gluon cross section is equivalent to finding the
classical field produced by two colliding nuclei or hadrons which was
found at the lowest order in \cite{claa}. The problem of inclusive
gluon production in pA including all orders of rescatterings in the
nucleus, i.e., resumming all powers of $\as^2 A^{1/3}$, has been
solved in \cite{kmu} and the result was later reproduced in
\cite{kop,dm}. At the same time the question of inclusive gluon 
production in DIS has not been studied yet. The calculation there is a
little different from pA and will be discussed later in this paper. In
AA collisions the problem is more complicated than the single nucleus
case of DIS since now one has to include rescatterings of the produced
gluon in the fields of nucleons in both nuclei. The gluon production
problem in AA is of utmost importance for initial conditions for
quark-gluon plasma (QGP) formation in heavy ion collisions. An
important recent progress has been made in \cite{kv,yuri1,KN,KLM} on
understanding the distribution of produced gluons in AA and in
\cite{bmss} on exploring the possible thermalization of gluons in the
saturation model. At the moment nobody found a way to include the
quantum corrections to the classical inclusive gluon production cross
section neither in DIS nor in pA or AA and this important question
still remains open.

The issue of multiple rescatterings in a medium produced in the heavy
ion collisions is related to the problem of energy loss of produced
particles \cite{gw}. Recently there has been developed a vigorous
activity in the field of energy loss \cite{bdmps,zak,urs} due to
sensitivity of this observable to the possible QGP formation in heavy
ion collisions. Multiple rescatterings in the cold nuclear medium
\cite{bdmps} are of similar nature to multiple rescatterings of the
saturation calculations \cite{kmu} and the interplay of two notions
could be very important for our understanding of nuclear collisions.

Exclusive observables, such as diffractive cross section and particle
production are also very important for our understanding of high
energy scattering. The total diffractive cross section and the
corresponding structure function have been found for DIS in the
quasi-classical approximation in \cite{Heb,km,kw}. A one-gluon
correction to the total diffractive cross section has been found in
\cite{kw}. The second stage of the saturation calculations has  
already been completed for diffraction: in \cite{kl} an equation has
been written resumming all leading logarithmic corrections (multiple
pomeron exchanges) to the total diffractive cross section in
DIS. Exclusive vector meson production including all the multiple
rescatterings at the quasi-classical level in DIS has been studied in
\cite{vec,kw}.

Nevertheless the question of calculating the diffractive gluon
production cross section in the quasi-classical approximation has not
been addressed yet in the literature. The problem is very important
for the proton-nucleus collision experiments at RHIC and for DIS at
HERA and we are going to study it in this paper. We begin in Sect. IIA
by considering the case of a quarkonium scattering on a target
nucleus. We are interested in the gluon production process where the
nucleus remains intact with no restrictions imposed on possible final
states of the $q\bar q$ pair.  The resulting gluon production cross
section is given by Eqs. (\ref{qqax}) and (\ref{qqaxk}) and includes
the effects of all multiple rescatterings in the background fields of
the nucleons in the nucleus in the quasi-classical approximation
(equivalent to resummation of powers of $\as^2 A^{1/3}$). We then
continue in Sect. IIB by applying the developed techniques to the case
of proton-nucleus collisions, where the proton is approximated as a
color single state of three valence quarks. The calculation for the pA
case is a little more involved than the calculation for
quarkonium-nucleus scattering. The answer in a more compact form is
given by Eqs. (\ref{MM}), (\ref{vv}), (\ref{ppi}), (\ref{pax}). We
argue that the presence of intrinsic sea quarks and gluons in the
projectile's wave function may tremendously complicate the calculation
of diffractive production cross sections, while posing no threat to
inclusive cross sections.  In Sect. III we generalize the result of
Sect. IIA to the case of deep inelastic scattering. We discuss the
differences between inclusive gluon production in DIS and pA, which
are rooted in the model of the proton one uses. We conclude the paper
by writing down an expression for inclusive gluon production in DIS
given by \eq{disxi2}.

\section{Hadron--Nucleus Collisions}

In this section we will first derive an expression for diffractive
gluon production cross section in the quarkonium--nucleus
scattering. We will continue by generalizing the result to the case of
diffractive proton--nucleus scattering.

\subsection{Quarkonium--Nucleus Collisions}

Let us start by considering a scattering of a $q \bar q$ (quarkonium)
state on a target nucleus. We want to calculate diffractive gluon
production cross section. That is we are interested in the
hadron-nucleus scattering processes where a soft (small-$x$) gluon is
produced in the central rapidity region while the nucleus remains
intact in the color neutral state. The diffractive gluon production
cross section could be written as a convolution of the onium's wave
function squared with the diffractive gluon production cross section
for the $q \bar q$ pair scattering on a nucleus
\be\label{x}
\frac{d \sigma^{onium A}_{diff}}{d^2k \, d y} \, = \, \int d^2 r \, d \alpha \,
|\Psi^{onium} ({\underline r},\alpha)|^2 \, \frac{d {\hat
\sigma}^{q{\bar q}A}_{diff}}{d^2k \, d y} ({\underline r}),
\ee
where ${\underline r}$ is the transverse separation of the quarks in
the onium and $\alpha$ is the light cone momentum fraction carried by
the quark.

\begin{figure}
\begin{center}
\epsfxsize=16.3cm
\leavevmode
\hbox{ \epsffile{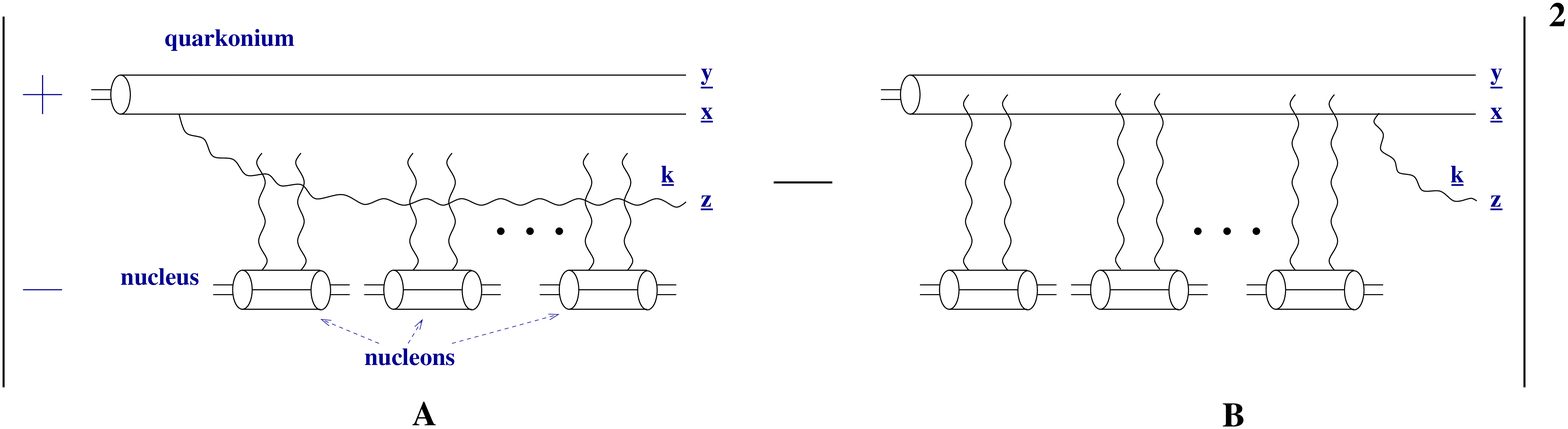}}
\end{center}
\caption{Diagrams contributing to the diffractive gluon production 
cross section in the quarkonium-nucleus scattering.}
\label{qqa}
\end{figure}

We are going to perform our calculations of $d {\hat
\sigma}^{q{\bar q}A}_{diff}/d^2k \, d y$ in the time-ordered light
cone perturbation theory \cite{bl} working in the light cone gauge of
the projectile hadron $A_+ = 0$ \cite{kmu}. Similarly to
\cite{kmu,dm,yuri1,km,kw} we distinguish the cases when the gluon is
present in the quarkonium's wave function before the collision and
when it is emitted after the collision. The diagrams contributing to
the diffractive gluon production are shown in \fig{qqa}. We assume
that the quark and the antiquark in the quarkonium are in the color
singlet state, that is there is no soft non-perturbative partons which
may carry some of the color in the quarkonium wave function making the
$q\bar q$ pair non-singlet in color.  The graph in
\fig{qqa}A represents the case when the $q{\bar q}$ state develops a
gluon fluctuation long before hitting the target. Subsequently quark,
antiquark and the gluon rescatter in the target nucleus leaving it
intact since we are interested in diffractive processes
\cite{Heb,km}. The exchanged gluons coming from the nucleons 
in the nucleus can connect to either quark, antiquark or gluon lines
and we have to sum over all of these possibilities. This is
represented in \fig{qqa} by not connecting gluon lines specifically to
any of the projectile parton lines. The diagram in \fig{qqa}B
corresponds to the case when only $q{\bar q}$ pair rescatters in the
nucleus and the gluon is emitted after the scattering. Again both the
quark and the antiquark lines interact with the nucleus which is
represented by the gluon lines not connecting to any specific quark
line. Finally in the diagrams of \fig{qqa} the gluon is shown to be
emitted off one of the quark lines only. To obtain the answer one of
course has to sum over all possible gluon emissions off the quark and
antiquark lines in the amplitude and in the complex conjugate
amplitude.

To calculate the diagrams in \fig{qqa} one needs to know forward
``propagators'' of the $q{\bar q}$ and $q{\bar q}G$ states through the
nucleus. The propagator of the quark-antiquark state through the
nucleus is well known and is given by \cite{Mue,kmu,yuri1,km,kw}
\be
\exp \left( - \frac{C_F}{4 \, N_c} \, 
({\underline x} - {\underline y})^2 \, Q_s^2 \right)
\ee
where ${\underline x}$ and ${\underline y}$ are the transverse
coordinates of the quark and the antiquark correspondingly and we have
used a shorthand notation \cite{bdmps,kmu,yuri1}
\be\label{xqs}
({\underline x} - {\underline y})^2 Q_s^2 \, = \, ({\underline x} -
{\underline y})^2 \ \frac{8 \pi^2 \as N_c \sqrt{R^2 - b^2}}{N^2_c - 1}
\, \rho \, xG (x, 1/|{\underline x} - {\underline y}|^2),
\ee
with $\rho = A / [(4/3) \pi R^3]$ the density of the atomic number $A$
and ${\underline b}$ the impact parameter of the $q{\bar q}$
pair. Here the nucleus is assumed to be a sphere of radius $R$, but
the result could be easily generalized to other geometries. The gluon
distribution function in \eq{xqs} at the two gluon level is given by
\cite{bdmps}
\be\label{xg}
xG (x, 1/{\underline x}^2) \ = \ \frac{\as C_F}{\pi} \, \ln
\frac{1}{{\underline x}^2 \mu^2},
\ee
with $\mu$ some infrared cutoff. In principle the {\it saturation
scale} $Q_s$ has to be found from the implicit equation
\be\label{qs}
Q_s^2 (b) \ = \ \frac{8 \pi^2 \as N_c \sqrt{R^2 - b^2}}{N^2_c - 1} \, \rho
\, xG (x, Q_s^2(b)).
\ee
Since the gluon distribution of \eq{xg} is a slowly (logarithmically)
varying function of transverse distance in certain problems the $Q_s$
dependence on the right hand side of \eq{qs} could be neglected
turning \eq{qs} from equation into an equality.

\begin{figure}
\begin{center}
\epsfxsize=12cm
\leavevmode
\hbox{ \epsffile{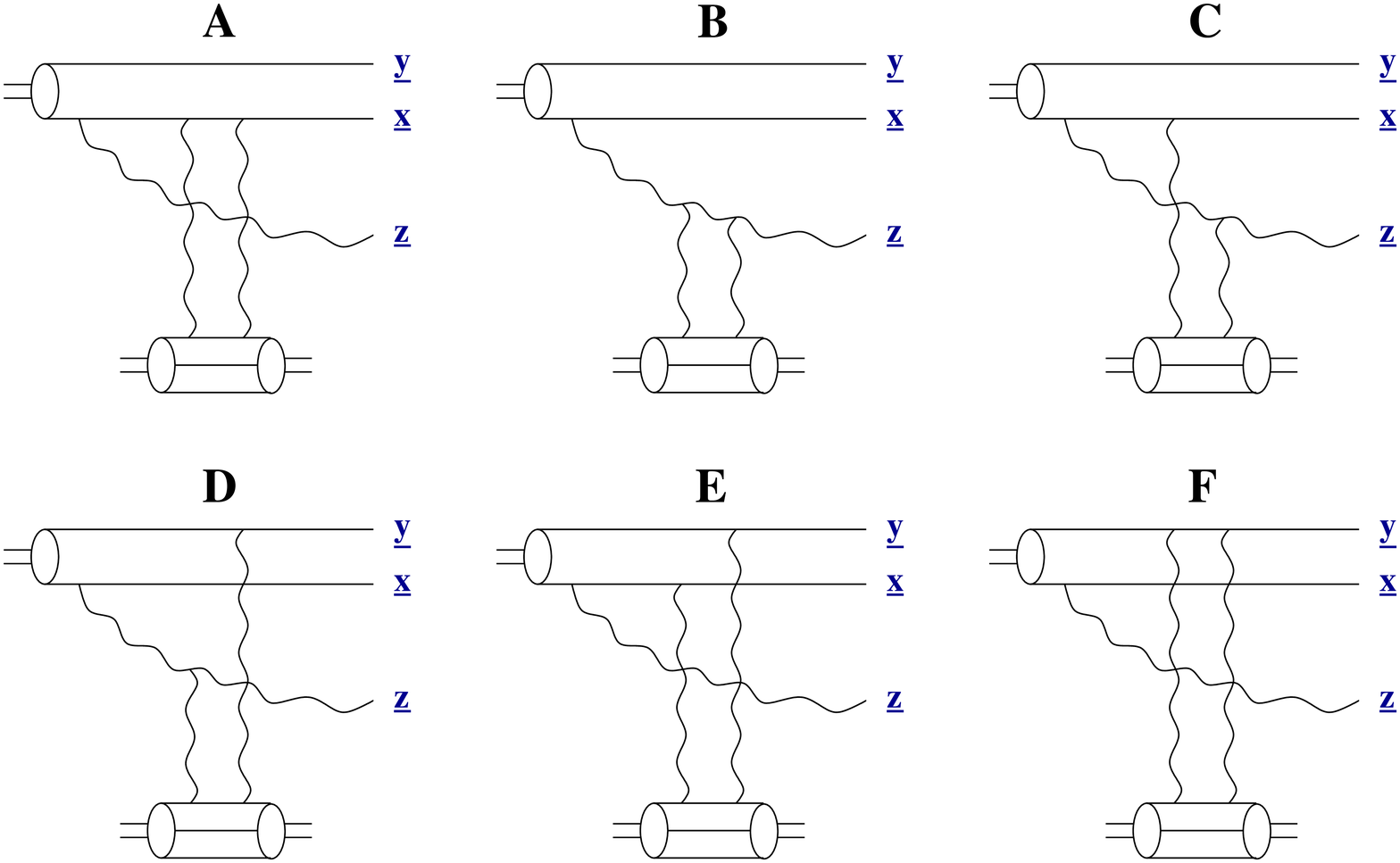}}
\end{center}
\caption{Diagrams included in a diffractive scattering of the $q{\bar q}G$ state on a 
single nucleon in the target nucleus.}
\label{qqg}
\end{figure}

The ``propagator'' of the $q{\bar q}G$ state through the nucleus is a
little harder to calculate. To do this let us consider the interaction
between the $q{\bar q}G$ state and the first nucleon in the nucleus
that exchanges gluons with it. The diagrams we need to sum are
depicted in \fig{qqg}. To write down the contributions of these
diagrams let us first introduce the gluon-nucleon interaction
potential as a Fourier transform into coordinate space of the
normalized gluon-nucleon scattering amplitude $\frac{1}{\sigma} \,
\frac{d \sigma}{d^2 l}$ \cite{bdmps,kmu}
\be\label{vt}
{\tilde V} ({\underline x}) \, = \, \int d^2 l e^{- i {\underline l}
\cdot {\underline x}} \, \frac{1}{\sigma} \, \frac{d \sigma}{d^2 l}.
\ee
With the help of \eq{vt} one can easily see that the contributions of
diagrams in \fig{qqg} are
\begin{mathletters}\label{qqgd}
\be
A \, = \, - \frac{C_F}{2 N_c} \, {\tilde V} ({\underline 0})
\ee
\be
B \, = \, - \frac{N_c}{2 N_c} \, {\tilde V} ({\underline 0})
\ee
\be
C \, = \, \frac{N_c}{2 N_c} \, {\tilde V} ({\underline z} - {\underline x})
\ee
\be
D \, = \, \frac{N_c}{2 N_c} \, {\tilde V} ({\underline z} - {\underline y})
\ee
\be
E \, = \, - \frac{1}{2 N_c^2} \, {\tilde V} ({\underline x} - {\underline y})
\ee
\be
F \, = \, - \frac{C_F}{2 N_c} \, {\tilde V} ({\underline 0})
\ee
\end{mathletters}
where we take a trace in the color space of the nucleons and average
over the colors of the nucleon, which yields a factor of $1/2 N_c$.

Summing up the diagrams of \fig{qqg} we obtain
\be\label{sumqqg}
A + B + \ldots + F \, = \, - \frac{1}{2} \, [{\tilde V} ({\underline
0}) - {\tilde V} ({\underline x} - {\underline z})] \, - \,
\frac{1}{2} \, [{\tilde V} ({\underline 0}) - {\tilde V} ({\underline
y} - {\underline z})] + \frac{1}{2 N_c^2} [{\tilde V} ({\underline 0})
- {\tilde V} ({\underline x} - {\underline y})] . 
\ee
The calculation which led to \eq{sumqqg} never assumed summation over
the colors of the quark-antiquark pair and the
gluon. Eqs. (\ref{qqgd}) explicitly show us that diffractive
scattering on a nucleon does not change the color structure of the
incoming $q{\bar q}G$ state. This allows us to easily see that the full
answer for the ``propagator'' of the $q {\bar q}G$ state through the
nucleus could be obtained by just exponentiating the interaction with
a single nucleon. Multiplying the right hand side of \eq{sumqqg} by $2
\sqrt{R^2 - b^2} \, \rho \, \sigma$ to take into account the density of the
nucleons in the nucleus \cite{kmu} yields
\be\label{pp}
- P ({\underline x}, {\underline y}, {\underline z}) \, \equiv \, 2
\sqrt{R^2 - b^2} \, \rho \, \sigma \, (A + B + \ldots + F) \, = \, - \frac{1}{8} 
\, ({\underline x} - {\underline z})^2 Q_s^2  - \frac{1}{8} 
\, ({\underline y} - {\underline z})^2 Q_s^2 + \frac{1}{8 N_c^2} 
\, ({\underline x} - {\underline y})^2 Q_s^2  .
\ee
Exponentiating \eq{pp} we obtain the propagator of the $q {\bar q}G$
state through the nucleus
\be\label{qqgp}
\exp \left(  - P ({\underline x}, {\underline y}, {\underline z}) \right) \, = \, 
\exp \left( - \frac{1}{8} 
\, ({\underline x} - {\underline z})^2 Q_s^2  - \frac{1}{8} 
\, ({\underline y} - {\underline z})^2 Q_s^2 + \frac{1}{8 N_c^2} 
\, ({\underline x} - {\underline y})^2 Q_s^2 \right),
\ee
which agrees with the result derived by Kopeliovich et al in
\cite{kop} and by Kovner and Wiedemann in \cite{kw}.

Now it is straightforward to write down the cross section for
diffractive gluon production. The cross section is given by the square
of the amplitude depicted in \fig{qqa} in momentum space. Fixing the
overall normalization yields
\ben
\frac{d {\hat \sigma}^{q{\bar q}A}_{diff}}{d^2k \, d y} \, = \,
 \frac{\as C_F}{\pi^2} \, 
\frac{1}{(2 \pi)^2} \, \int \, d^2 b \, d^2 z_1 \, d^2 z_2 \, 
e^{- i {\underline k} \cdot ({\underline z}_1 - {\underline z}_2)} \,
\left( \frac{{\underline z}_1- {\underline x}}{|{\underline z}_1- 
{\underline x}|^2} - \frac{{\underline z}_1- {\underline
y}}{|{\underline z}_1- {\underline y}|^2}\right) \cdot \left(
\frac{{\underline z}_2- {\underline x}}{|{\underline z}_2- {\underline
x}|^2} - \frac{{\underline z}_2- {\underline y}}{|{\underline z}_2-
{\underline y}|^2}\right)
\een
\be\label{qqax}
\times \left( e^{- P ({\underline x}, {\underline y}, {\underline z}_1)} - 
e^{ - \frac{C_F}{4 \, N_c} \, ({\underline x} - {\underline y})^2 \,
Q_s^2}
\right) \, \left( e^{- P ({\underline x}, {\underline y}, {\underline
z}_2)} - e^{ - \frac{C_F}{4 \, N_c} \, ({\underline x} - {\underline y})^2 \,
Q_s^2} \right),
\ee
where $z_1$ and $z_2$ are the transverse coordinates of the gluon in
the amplitude and in the complex conjugate amplitude
correspondingly. \eq{qqax} gives us the cross section of diffractive
gluon production in quarkonium-nucleus collisions. In general the
integrals over $z_1$ and $z_2$ in \eq{qqax} are very hard to do
analytically and they should probably be done numerically. However,
for not very large transverse momentum of the produced gluon, $k_\perp
\lsim Q_s$, one can neglect the logarithmic dependence on
$|{\underline x} - {\underline y}|$ of the gluon distribution in
\eq{xqs} making analytical evaluation of the integral in \eq{qqax}
possible \cite{kmu,yuri1}. To calculate the integrals in
\eq{qqax} in this logarithmic approximation we need to estimate the
following integral
\be\label{i1}
I ({\underline k}, {\underline x}; {\underline y}) \, = \, \int \, d^2 z \, e^{- i
{\underline k} \cdot {\underline z}} \, \frac{{\underline z} -
{\underline x}}{|{\underline z} - {\underline x}|^2} \, e^{-
\frac{1}{8} \, ({\underline z} - {\underline x})^2 Q_s^2 - \frac{1}{8}
\, ({\underline z} - {\underline y})^2 Q_s^2}.
\ee
After redefinition of variables ${\underline z} \rightarrow
{\underline z} + {\underline x}$ the integral of \eq{i1} could be
rewritten as
\be\label{i2}
I ({\underline k}, {\underline x}; {\underline y}) \, = \, i e^{- i {\underline k}
\cdot {\underline x} - \frac{1}{8} \, ({\underline x} - {\underline
y})^2 Q_s^2} \nabla_k \int_0^\infty \, \frac{d z}{z} \, e^{-
\frac{1}{4} z^2 Q_s^2} \, \int_0^{2 \pi} \, d \phi \, e^{- i k z \cos \phi 
- \frac{1}{4} z |{\underline x} - {\underline y}| Q_s^2 \cos (\phi - \beta)}
\ee
with $\phi$ the angle between ${\underline z}$ and ${\underline k}$
and $\beta$ the angle between ${\underline z}$ and ${\underline x} -
{\underline y}$. The integration over $\phi$ in \eq{i2} could be done
(see ${\bf 3.937}.2$ in \cite{gr}) yielding
\be\label{i3}
I ({\underline k}, {\underline x}; {\underline y}) \, = \, 2 \pi i
e^{- i {\underline k}
\cdot {\underline x} - \frac{1}{8} \, ({\underline x} - {\underline
y})^2 Q_s^2} \nabla_k \int_0^\infty \, \frac{d z}{z} \, e^{-
\frac{1}{4} z^2 Q_s^2} \, I_0 (z \sqrt{- {\underline \lambda}^2}),
\ee
where we have introduced a two dimensional complex vector
\be\label{lam}
{\underline \lambda} \, = \, {\underline k} - \frac{i}{4} \, Q_s^2 \,
({\underline x} - {\underline y}).
\ee
Differentiating \eq{i3} over $k$ and integrating over $z$ we finally
obtain
\be\label{i4}
I ({\underline k}, {\underline x}; {\underline y}) \, = \, - 2 \pi i
\, \frac{{\underline
\lambda}}{{\underline \lambda}^2} \, \left( 1 - e^{- {\underline \lambda}^2/Q_s^2} 
\right) \, e^{- i {\underline k}
\cdot {\underline x} - \frac{1}{8} \, ({\underline x} - {\underline
y})^2 Q_s^2}.
\ee
Inserting \eq{pp} into \eq{qqax} and employing \eq{i4} yields
\ben
\frac{d {\hat \sigma}^{q{\bar q}A}_{diff}}{d^2k \, d y} \, = \,
 \frac{\as C_F}{\pi^2} \, \int \, d^2 b \, \left|  \frac{{\underline
\lambda}}{{\underline \lambda}^2} \, \left(  1 - e^{- {\underline 
\lambda}^2/Q_s^2} \right) \, e^{- i {\underline k}
\cdot {\underline x}} -   \frac{{\underline
\lambda}^*}{{\underline \lambda}^{*2}} \, \left(  1 - e^{- {\underline 
\lambda}^{*2}/Q_s^2} \right) \, e^{- i {\underline k}
\cdot {\underline y}}  \right.
\een
\be\label{qqaxk}
- \left. \frac{{\underline k}}{{\underline k}^2} \, 
\left(  e^{- i {\underline k}
\cdot {\underline x}} - e^{- i {\underline k}
\cdot {\underline y}} \right) \right|^2 \, e^{- \frac{C_F}{2 \, N_c} 
({\underline x} - {\underline y})^2 Q_s^2} ,
\ee
where ${\underline \lambda}^*$ is the complex conjugate of
${\underline \lambda}$.

\eq{qqaxk} together with \eq{lam} provides us with the expression for 
the diffractive gluon production cross section for quarkonium-nucleus
collisions explicitly in terms of gluon's transverse momentum
${\underline k}$ and the transverse separation of the onium state
${\underline x} - {\underline y}$. After integration over angles
between ${\underline k}$ and ${\underline x} - {\underline y}$
\eq{qqaxk} could be easily rewritten in terms of the invariant mass of
the produced particles which is approximately equal to $M_X^2 \approx
{\underline k}^2 / x$ where $y = \ln 1/x$.

To the degree that a quarkonium state can serve as a model of a baryon
projectile such as proton, Eqs. (\ref{qqax}) and (\ref{qqaxk}) could
describe the diffractive gluon production cross section in $pA$
collisions. However, a much better model of $pA$ collisions would
involve a realistic proton consisting of three valence quarks and we
are going to address this problem in Sect. IIB.

\subsection{Proton--Nucleus Collisions}

Diffractive gluon production in proton-nucleus collisions is different
from our previous discussion of onium-nucleus scattering only by the
fact that the proton consists of three valence quarks instead of 
quark-antiquark pair. Again we assume that the three quarks in the
proton are in a color singlet state by themselves, i.e., there is no
soft non-perturbative gluons in the proton's wave function to modify
this picture.

Similarly to the quarkonium case diffractive gluon production cross
section could be expressed as a convolution of the proton's wave
function and the diffractive cross section of a $qqq$ state on the
target nucleus:
\be\label{px}
\frac{d \sigma^{pA}_{diff}}{d^2k \, d y} \, = \, \int d^2 x_{13} \, 
d^2 x_{23} \, d \alpha_1 \, d \alpha_2 \, |\Psi^p ({\underline
x}_{13}, {\underline x}_{23},
\alpha_1, \alpha_2 )|^2 \, \frac{d {\hat
\sigma}^{qqqA}_{diff}}{d^2k \, d y} ({\underline x}_{13}, {\underline x}_{23}),
\ee
with ${\underline x}_{13}, {\underline x}_{23}$ the distances between
valence quarks in the proton and $\alpha_1, \alpha_2$ light cone
fractions of the proton's momentum carried by two of the quarks.

\begin{figure}
\begin{center}
\epsfxsize=16.3cm
\leavevmode
\hbox{ \epsffile{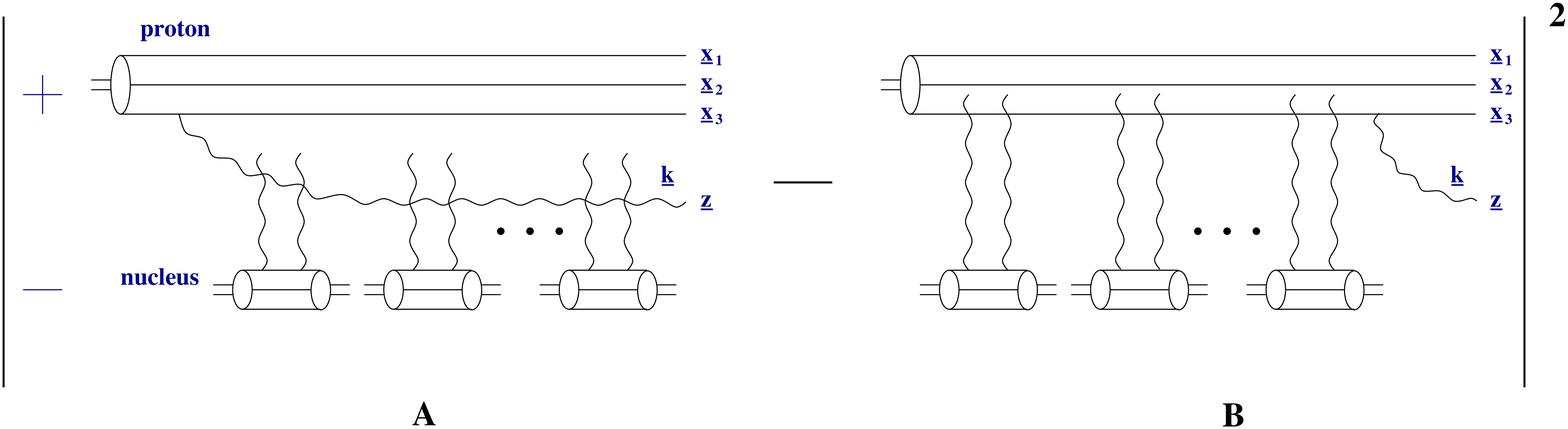}}
\end{center}
\caption{Diffractive gluon production in proton-nucleus scattering (pA).}
\label{pa}
\end{figure}

The diagrams describing the diffractive gluon production in pA are
shown in \fig{pa}. They are similar to the onium-nucleus scattering
diagrams of \fig{qqa} with three quarks instead of $q\bar q$ pair in
the projectile. The diagrams in \fig{pa} demonstrate that in order to
calculate diffractive gluon production cross section we need to know
the diffractive ``propagators'' of the $qqq$ and $qqqG$ states through
the nucleus.

\begin{figure}
\begin{center}
\epsfxsize=9cm
\leavevmode
\hbox{ \epsffile{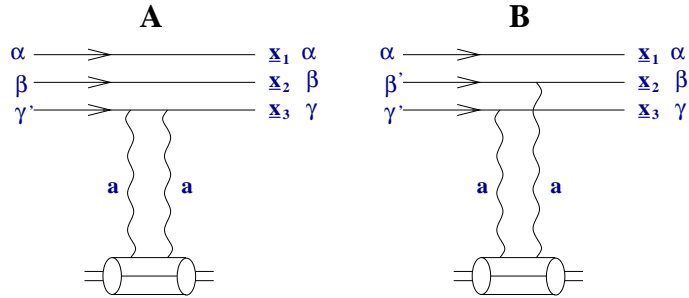}}
\end{center}
\caption{Diagrams contributing to the interaction of the proton with 
a nucleon in the target nucleus.}
\label{qqq}
\end{figure}

First we are going to calculate the ``propagator'' of the $qqq$ state
through the nucleus, which is needed for the diagram in
\fig{pa}B. Similar to what was done in Sect. IIA we need to consider
scattering of three quarks on a single nucleon. The contributing
diagrams are shown in \fig{qqq}. There are two types of diagrams in
proton-nucleon interaction: there are diagrams where both gluons
connect to the same quark line, as shown in \fig{qqq}A and there are
diagrams where gluons connect to different quark lines, as shown in
\fig{qqq}B. The color singlet configuration of three quarks is achieved 
by antisymmetrization over their color indices with the help of
$\epsilon_{\alpha \beta \gamma}$.

The diagram in \fig{qqq}A easily yields 
\be\label{aa}
{\cal A} \, = \, - \frac{C_F}{2 N_c} \, {\tilde V} ({\underline 0}).
\ee
The diagram in \fig{qqq}B is a little harder to calculate. The color
factor there is given by 
\ben
\frac{1}{2 N_c} \, \epsilon_{\alpha \beta' \gamma'} \, (T^a)_{\beta \beta'} 
\, (T^a)_{\gamma \gamma'}.
\een
Using the Fierz identity
\be\label{fierz}
(T^a)_{\alpha\beta} \, (T^a)_{\gamma\delta} \, = \, \frac{1}{2} \,
\delta_{\alpha\delta} \, \delta_{\beta\gamma} - \, \frac{1}{2 N_c} \, 
\delta_{\alpha\beta} \, \delta_{\gamma\delta}
\ee
we readily evaluate the diagram in \fig{qqq}B to be
\be\label{bb}
{\cal B} \, = \, \frac{N_c + 1}{(2 N_c)^2} \, {\tilde V} ({\underline
x}_3 - {\underline x}_2).
\ee
To obtain the proton's propagator out of Eqs. (\ref{aa}) and
(\ref{bb}) we have to sum over gluons' connections to all valence
quarks in the proton and multiply the result by $2 \sqrt{R^2 - b^2}
\, \rho \, \sigma$. The final expression for the proton's propagator is
\be\label{pprop}
\exp \left( - \frac{N_c + 1}{4 \, (2 N_c)^2} \, \sum_{i,j=1; i \neq j}^{N_c} 
\, ({\underline x}_i - {\underline x}_j)^2 Q_s^2 \right)
\ee
which for the case of three colors becomes
\be\label{pprop3}
\exp \left( - \frac{1}{18} \, \left[ ({\underline x}_1 - {\underline x}_2)^2 Q_s^2 
+ ({\underline x}_2 - {\underline x}_3)^2 Q_s^2 
+ ({\underline x}_1 - {\underline x}_3)^2 Q_s^2 \right] \right).
\ee

\begin{figure}
\begin{center}
\epsfxsize=9cm
\leavevmode
\hbox{ \epsffile{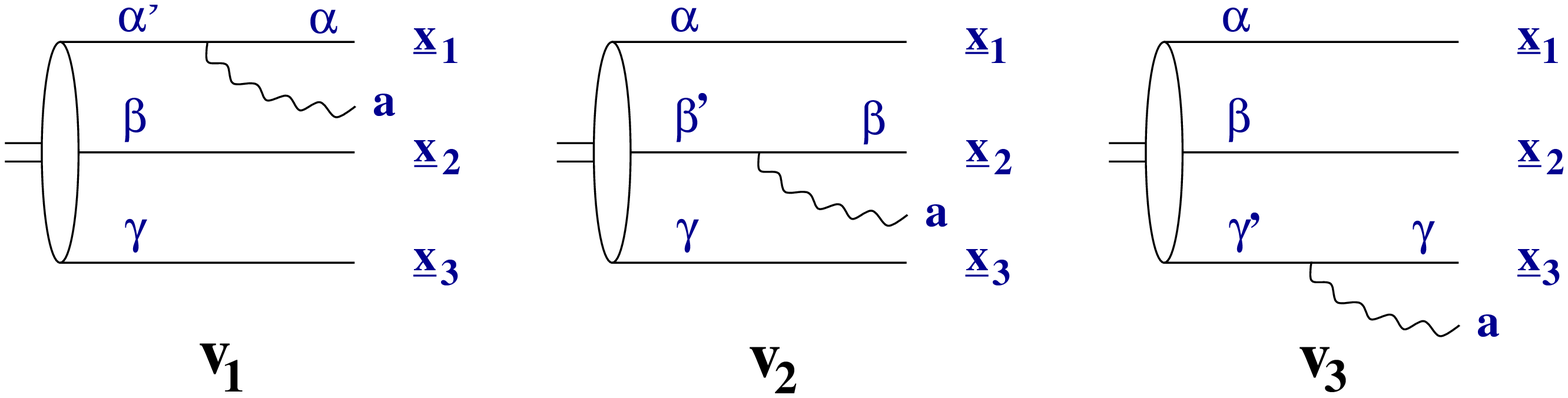}}
\end{center}
\caption{Three possible ways of generating a $qqqG$ fluctuation in a proton.}
\label{qqqg}
\end{figure}

Calculation of the $qqqG$ ``propagator'' in the nucleus is more
complicated because, unlike the above cases, the scattering of a
single nucleon is not diagonal in the color space. Thus instead of
exponentiating a scalar we will have to exponentiate a matrix in the
color space. (For similar calculations of different quantities see
\cite{kw}.) To determine the color matrix we have to again consider 
the scattering of the $qqqG$ system on a single nucleon. First let us
note that the $qqqG$ system can arise due to gluon emissions off each
of the valence quarks, giving rise to a different color structure in
each case. The possibilities are outlined in \fig{qqqg}. Quarks there
are labeled $1,2,3$ carrying color indices $\alpha,\beta,\gamma$
correspondingly.

The three states depicted in \fig{qqqg} correspond to three color
matrices, which we will label $v_1$, $v_2$ and $v_3$ such that
\be\label{vs}
v_1 \, = \, \frac{1}{\sqrt{6}} \, \epsilon_{\alpha'\beta\gamma} \,
(T^a)_{\alpha\alpha'}, \hspace*{1cm} v_2 \, = \, \frac{1}{\sqrt{6}} \,
\epsilon_{\alpha\beta'\gamma} \, (T^a)_{\beta\beta'}, \hspace*{1cm}
v_3 \, = \, \frac{1}{\sqrt{6}} \, \epsilon_{\alpha\beta\gamma'} \,
(T^a)_{\gamma\gamma'}.
\ee
The factor of $1/\sqrt{6}$ is arises in \eq{vs} due to
$(\epsilon_{\alpha\beta\gamma})^2 = N_c ! = 6$ and is included to
factor out the color structure of the proton's wave function.  

As we will see below interactions with nucleons only transform each of
the states in \eq{vs} into linear combinations of the others. It is
therefore convenient to make an orthonormal basis out of these
matrices. We note that since due to $SU(N_c)$ group properties
\be
v_1 + v_2 + v_3 = 0
\ee
the matrices of \eq{vs} are not linearly independent. (In fact they
represent the roots of $SU(N_c)$.) We are going to choose the
following orthonormal basis
\be\label{us}
u_2 = v_2 + \frac{1}{2} \, v_3 , \hspace*{1cm} u_3 = \frac{\sqrt{3}}{2} \, v_3,
\ee
where we have explicitly inserted $N_c = 3$ and $C_F = 4/3$. Since in
\fig{qqqg} we explicitly consider a baryon (proton) consisting of three 
valence quarks our results from here to the end of the section will be
derived for $N_c = 3$. To generalize this discussion to an arbitrary
$N_c$ one would have to introduce $N_c$ different matrices $v_1,
\ldots, v_{N_c}$ and work in the $N_c -1$ dimensional space of the
orthogonal basis $u_2, \ldots , u_{N_c}$. Doing that is beyond the
scope of this paper.

\begin{figure}
\begin{center}
\epsfxsize=16.3cm
\leavevmode
\hbox{ \epsffile{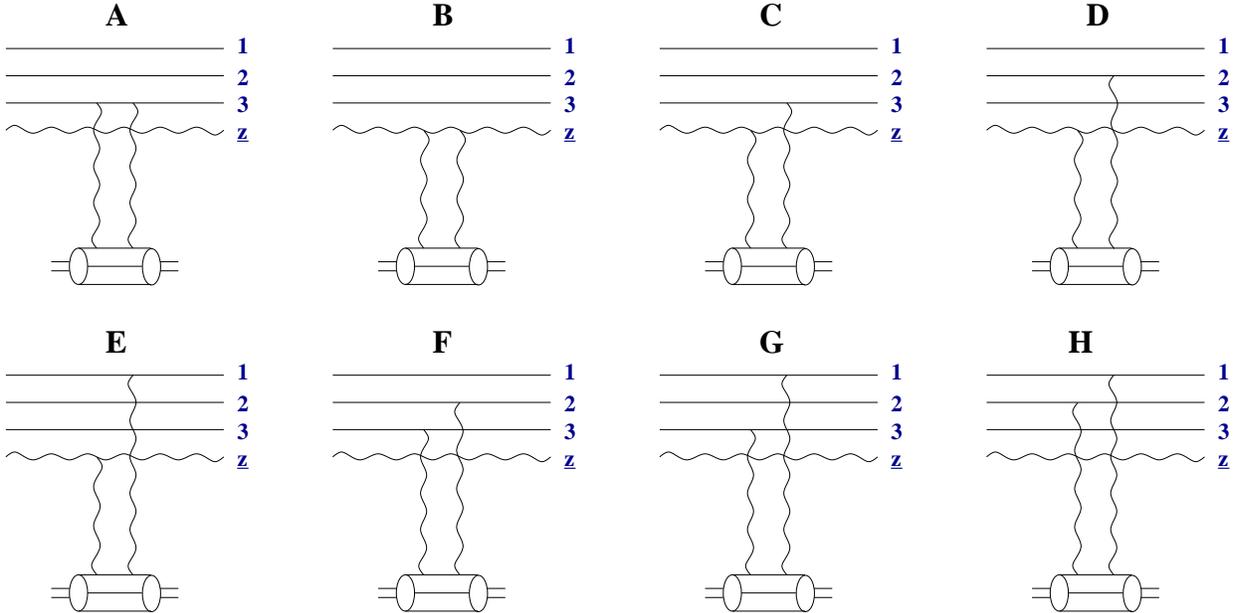}}
\end{center}
\caption{Interaction of the $qqqG$ state with a nucleon.}
\label{qqqgp}
\end{figure}

Let us now consider interaction of the $qqqG$ state with a single
nucleon in the nucleus. The corresponding diagrams are depicted in
\fig{qqqgp}. Each of the diagrams is a matrix in the $u_2, u_3$ linear
space. To calculate the diagrams of \fig{qqqgp} we have to first
calculate the action of each of them on $v_2$ and $v_3$ and then use
\eq{us} to rewrite their contributions in the $(u_2, u_3)$ basis. The
result of a rather tedious calculation yields
\begin{mathletters}\label{ms}
\be\label{aaa}
M_A \, = \, - \frac{2}{3} \, {\tilde V} ({\underline 0}) \pmatrix{1 & 0 \cr 0 & 1}
\ee
\be
M_B \, = \, - \frac{1}{2} \, {\tilde V} ({\underline 0}) \pmatrix{1 & 0 \cr 0 & 1}
\ee
\be
M_C \, = \, \frac{1}{2} \, {\tilde V} ({\underline z} - {\underline x}_3) 
\pmatrix{1/3 & 0 \cr 0 & 1}
\ee
\be
M_D \, = \, \frac{1}{4} \, {\tilde V} ({\underline z} - {\underline x}_2) 
\pmatrix{5/3 & - 1/\sqrt{3} \cr - 1/\sqrt{3}  & 1}
\ee
\be
M_E \, = \, \frac{1}{4} \, {\tilde V} ({\underline z} - {\underline x}_1) 
\pmatrix{5/3 & 1/\sqrt{3} \cr 1/\sqrt{3}  & 1}
\ee
\be
M_F \, = \, \frac{1}{4} \, {\tilde V} ({\underline x}_2 - {\underline x}_3) 
\pmatrix{5/9 & 1/\sqrt{3} \cr 1/\sqrt{3}  & -1/9}
\ee
\be
M_G \, = \, \frac{1}{4} \, {\tilde V} ({\underline x}_1 - {\underline x}_3) 
\pmatrix{5/9 & - 1/\sqrt{3} \cr - 1/\sqrt{3}  & -1/9}
\ee
\be
M_H \, = \, {\tilde V} ({\underline x}_1 - {\underline x}_2) 
\pmatrix{-1/9 & 0 \cr 0  & 2/9}
\ee
\end{mathletters}
where in \eq{aaa} we summed over gluon connections to all three
valence quarks. Summing up the contributions of \eq{ms} and
multiplying them by $2 \sqrt{R^2 - b^2} \, \rho \, \sigma$ yields
\be\label{M}
- M ({\underline z}) \, \equiv \, 2 \sqrt{R^2 - b^2} \, \rho \, \sigma \,
(M_A + \ldots + M_H)
\ee
with the matrix $M ({\underline z})$ given by
\be\label{MM}
M ({\underline z}) \, = \, \pmatrix{\frac{1}{6} \, \zeta_3 +
\frac{5}{12} \, (\zeta_2 + \zeta_1) + \frac{5}{36} \, (\chi_{23} +
\chi_{13}) - \frac{1}{9} \, \chi_{12} & \frac{1}{4 \sqrt{3}} \, ( -
\zeta_2 + \zeta_1 + \chi_{23} - \chi_{13}) \cr \frac{1}{4 \sqrt{3}} \,
( - \zeta_2 + \zeta_1 + \chi_{23} - \chi_{13}) & \frac{1}{2} \,
\zeta_3 + \frac{1}{4} \,
( \zeta_2 + \zeta_1) - \frac{1}{36} \, (\chi_{23} + \chi_{13}) +
\frac{2}{9} \, \chi_{12}}
\ee
where we have defined
\be
\zeta_i \, \equiv \, \frac{1}{4} \, ({\underline z} - {\underline x}_i)^2 Q_s^2
\ee
and
\be
\chi_{ij} \, \equiv \, \frac{1}{4} \, ({\underline x}_i - {\underline x}_j)^2 Q_s^2. 
\ee
The ``propagator'' of the $qqqG$ state through the nucleus is given by 
\be
\exp [ - M ({\underline z})],
\ee
which is also a matrix. 

Now we are in a position to write down a compact expression for the
diffractive gluon production cross section in the proton-nucleus
collisions. The initial proton state in the amplitude and in the
complex conjugate amplitude could be in either one of the three states,
$v_1$, $v_2$ and $v_3$, which are given by
\be\label{vv}
v_1^T \, = \, ( - 1 \hspace*{2mm} , \hspace*{3mm} - \frac{1}{\sqrt{3}}), 
\hspace*{1cm} v_2^T \, = \, (1 \hspace*{2mm} , \hspace*{3mm} - \frac{1}
{\sqrt{3}}), \hspace*{1cm} v_3^T \, = \, (0 \hspace*{2mm} ,
\hspace*{3mm} \frac{2}{\sqrt{3}})
\ee 
in the $(u_2,u_3)$ basis. If the proton in the amplitude is in the
state $v_i$ and in the complex conjugate amplitude it is in the state
$v_j$ we can define a diffractive overlap functions
\be\label{ppi}
\Pi_{ij} \, \equiv \, v_i^T \left( e^{- M({\underline z}_1)} - e^{-(2/9) 
(\chi_{12} + \chi_{13} + \chi_{23})} \right) \left( e^{- M({\underline
z}_2)} - e^{-(2/9) (\chi_{12} + \chi_{13} + \chi_{23})} \right) v_j,
\ee
where, as in Sect. IIA , ${\underline z}_1$ and ${\underline z}_2$ are
the gluon's transverse coordinates in the amplitude and in the complex
conjugate amplitude. The second exponent in \eq{ppi} comes from
\fig{pa}B with the propagator of the proton in the nucleus given by
\eq{pprop3}, which is a unit matrix in the $(u_2,u_3)$ color space. 
The diffractive gluon production cross section in pA can be written in
terms of $\Pi_{ij}$'s as
\be\label{pax}
\frac{d {\hat \sigma}^{qqqA}_{diff}}{d^2k \, d y} \, = \,
 \frac{\as}{(2 \pi)^2 \, \pi^2} \, \int \, d^2 b \, d^2 z_1 \, d^2 z_2
 \, \sum_{i=1}^3 \sum_{j=1}^3 \frac{{\underline z}_1 - {\underline
 x}_i}{|{\underline z}_1 - {\underline x}_i|^2} \cdot \frac{{\underline
 z}_2 - {\underline x}_j}{|{\underline z}_2 - {\underline x}_j|^2} \, \Pi_{ij}.
\ee
Eqs. (\ref{MM}), (\ref{vv}), (\ref{ppi}), (\ref{pax}), together with
\eq{px} provide us the answer for diffractive gluon production
cross section in pA collisions. The matrix expression for $\Pi_{ij}$
in \eq{ppi} could be rewritten in the usual ``scalar'' form. After
some lengthy algebra one arrives at
\ben
\Pi_{33} \, = \, \frac{4}{3} \, \left[ 4 \frac{M_{12} ({\underline z}_1) 
M_{12} ({\underline z}_2) }{\sqrt{D ({\underline z}_1) \, D ({\underline
z}_2) }} \, e^{- \frac{1}{2} \, [M_{11} ({\underline z}_1) + M_{22}
({\underline z}_1) + M_{11} ({\underline z}_2) + M_{22} ({\underline
z}_2)] } \, \sinh \frac{\sqrt{D ({\underline z}_1) }}{2} \, \sinh
\frac{\sqrt{D ({\underline z}_2) }}{2} \right.
\een
\be\label{p33}
 + \left( e^{- \frac{1}{2} \, [M_{11} ({\underline z}_1) + M_{22}
 ({\underline z}_1)]} \, R ({\underline z}_1) - e^{- \frac{2}{9} \,
 (\chi_{12} + \chi_{23} + \chi_{13})} \right)
\left. \left( e^{- \frac{1}{2} \, [M_{11} ({\underline z}_2) + M_{22}
({\underline z}_2)]} \, R ({\underline z}_2) - e^{- \frac{2}{9} \,
(\chi_{12} + \chi_{23} + \chi_{13})} \right) \right]
\ee
and
\ben
\Pi_{23} \, = \, - \frac{4}{\sqrt{3}} \, 
\frac{M_{12} ({\underline z}_2)}{\sqrt{D ({\underline z}_2)}} \, 
e^{- \frac{1}{2} \, [M_{11} ({\underline z}_2) + M_{22} ({\underline
z}_2)]} \, \sinh \frac{\sqrt{D ({\underline z}_2)}}{2} 
\een
\ben
\times \left( e^{-
\frac{1}{2} \, [M_{11} ({\underline z}_1) + M_{22} ({\underline
z}_1)]} \, L ({\underline z}_1) - e^{- \frac{2}{9} \, (\chi_{12} +
\chi_{23} + \chi_{13})} 
+ \frac{2}{\sqrt{3}} \, \frac{M_{12} ({\underline z}_1)}{\sqrt{D
({\underline z}_1)}} \, e^{- \frac{1}{2} \, [M_{11} ({\underline z}_1)
+ M_{22} ({\underline z}_1)]} \, \sinh \frac{\sqrt{D ({\underline
z}_1)}}{2} \right) 
\een
\ben
- \frac{2}{\sqrt{3}} \, \left( e^{- \frac{1}{2} \, [M_{11}
({\underline z}_2) + M_{22} ({\underline z}_2)]} \, R ({\underline
z}_2) - e^{- \frac{2}{9} \, (\chi_{12} +
\chi_{23} + \chi_{13})} \right) \, 
\left( \frac{2 \, M_{12} ({\underline z}_1)}{\sqrt{D ({\underline z}_1)}} 
\, e^{- \frac{1}{2} \, [M_{11} ({\underline z}_1) + M_{22} ({\underline z}_1)]} 
\, \sinh \frac{\sqrt{D ({\underline z}_1) }}{2} \right.
\een
\be\label{p23}
- \left. \frac{1}{\sqrt{3}} \, e^{- \frac{2}{9} \, (\chi_{12} +
\chi_{23} + \chi_{13})} +
\frac{1}{\sqrt{3}} \, e^{- \frac{1}{2} \, [M_{11} ({\underline z}_1) + M_{22} 
({\underline z}_1)]} \, R ({\underline z}_1) \right),
\ee
where we have introduced
\be\label{D}
D ({\underline z}) \, \equiv \, 4 \, [M_{12} ({\underline z})]^2 +
[M_{11} ({\underline z}) - M_{22} ({\underline z})]^2,
\ee
\be
R ({\underline z}) \, \equiv \, \cosh \frac{\sqrt{D ({\underline z}) }}{2}
+ \frac{M_{11} ({\underline z}) - M_{22} ({\underline z})}{\sqrt{D
({\underline z}) }} \, \sinh \frac{\sqrt{D ({\underline z}) }}{2} , 
\ee
\be
L ({\underline z}) \, \equiv \, \cosh \frac{\sqrt{D ({\underline z}) }}{2}
- \frac{M_{11} ({\underline z}) - M_{22} ({\underline z})}{\sqrt{D
({\underline z}) }} \, \sinh \frac{\sqrt{D ({\underline z}) }}{2} ,
\ee
and $M_{11} ({\underline z})$, $M_{22} ({\underline z})$ and $M_{12}
({\underline z})$ are the components of (symmetric) matrix $M
({\underline z})$ given in \eq{MM}. All other diagonal ($\Pi_{ii}$)
and off-diagonal ($\Pi_{ij}$, $i \neq j$) propagators could be
obtained from Eqs. (\ref{p33}) and (\ref{p23}) correspondingly by
appropriate permutations of the indices of $\zeta$'s and $\chi$'s.

We have calculated above the diffractive gluon production cross
section for quarkonium-nucleus and proton-nucleus collisions in the
quasi-classical approximation (Eqs. (\ref{qqax}) and (\ref{pax})). The
calculation for the case of pA collisions is much more sophisticated
than the calculation for onium-nucleus scattering. This is due to the
fact that the object in question is the diffractive cross section and
therefore interactions with spectator quarks have to be included. When
one calculates the total inclusive gluon production cross section the
interactions of the target nucleus with the spectator quarks (or
gluons) in the projectile disappear due to real-virtual cancellations
\cite{kmu,kop,dm,kw}. All one has to do there to obtain the answer is to 
calculate the interaction of a single projectile quark with the
nucleus and then convolute it with the quark distribution function in
the proton \cite{kmu,kop,dm,kw}\footnote{This statement depends on the
model of the proton: if we assume as above that the three valence
quarks are initially in the color singlet state then the produced
gluon could be emitted off different quarks in the amplitude and in
the complex conjugate amplitude (see Sect. III). Nevertheless
interactions with the quarks off which the gluon was {\it not} emitted
cancel.}. This is not the case for diffractive scattering: since the
final state here is not ``inclusive'', i.e., we imposed a rapidity gap
condition on the final state, only virtual exchanges contribute to
interaction with the target, as seen in Figs. \ref{qqa} and
\ref{pa}. Therefore there is no real-virtual cancellation for
interactions with spectator quarks (quarks off which the gluon was
{\it not} emitted neither in the amplitude nor in the complex
conjugate amplitude). Interactions with all partons in the proton's or
onium's wave function have to be included. That way, since our results
of Eqs. (\ref{qqax}) and (\ref{pax}) were derived for the models of
hadrons consisting of valence quarks one should use them to describe
the experimental data with caution. Realistic hadron's wave function
at high energy may include non-perturbative fluctuations producing the
so-called ``intrinsic'' sea quarks advocated by Brodsky et al in
\cite{brod} and quantum fluctuations due to perturbative QCD evolution
producing (``extrinsic'') sea quarks and gluons. Inclusion of either
one of the two effects would tremendously complicate the diffractive
gluon production cross section calculations presented above. Inclusion
of the effects of sea quarks and gluons in diffractive production
cross section is an important question which has to be addresses in a
separate study. At the moment we may argue that perturbatively
generated sea quarks and gluons would bring in higher powers of the
strong coupling constant $\as$ and could be neglected in the
quasi-classical approach employed here.

\section{Deep Inelastic Scattering}

We can generalize the results of Sect. IIA to the case of diffractive
gluon production in the deep inelastic scattering (DIS). The diagrams
important for the small-$x$ gluon production are the same diagrams as
shown in \fig{qqa} only with the wave function of a virtual photon
splitting into $q\bar q$ pair instead of onium wave function. The
expression for the diffractive gluon production cross section in DIS
is
\be\label{disdx}
\frac{d \sigma^{\gamma^* A \rightarrow q{\bar q} G A}_{diff}}{d^2 k \, dy} \, = \, 
\frac{1}{2 \pi^2} \, \int \, d^2 r \, d z \, \Phi^{\gamma^* \rightarrow q\bar q} 
({\underline r}, z) \, \frac{d {\hat \sigma}^{q{\bar q}A}_{diff}}{d^2
k \, dy}({\underline r}),
\ee
where $d {\hat \sigma}^{q{\bar q}A}_{diff} / d^2 k \, dy$ is given by
\eq{qqax}, ${\underline r} = {\underline x} - {\underline y}$ is the 
transverse separation of the quark-antiquark pair and $z$ is the
virtual photon's light cone momentum fraction carried by the
quark. The square of the virtual photon's wave function
$\Phi^{\gamma^* \rightarrow q\bar q} ({\underline r}, z)$ is a well
known function and could be found for instance in \cite{km}.

An interesting question to address is the calculation of the inclusive
gluon production cross section in DIS. This corresponds to the gluon
production cross section without any restrictions on the final state
of the target proton or nucleus. The problem is somewhat different
from the gluon production in pA as discussed in
\cite{kmu,kop,kw}. In the model of the proton considered in \cite{kmu,kop,kw}
the interacting valence quark in the proton could have an arbitrary
color in the initial state due to the soft intrinsic partons which
would randomize the colors of valence quarks. Thus an independent
summation was performed over the colors of this interacting
quark. This could not be done in DIS: due to perturbative nature of
the $q\bar q$ state at the leading order in $\as$ there is no quantum
fluctuations in the $q\bar q$ wave function randomizing the colors of
the quark and the antiquark. Instead, in quasi-classical approximation
employed here the quark and antiquark are in the color singlet state
and an independent summation over the colors of one of them is not
possible. Therefore we have to consider the diagrams with the gluon
emitted off both the quark and the antiquark.

There are two types of diagrams contributing to inclusive gluon
production cross section in DIS. The diagrams may have the gluon
emitted off the quark (antiquark) both in the amplitude and in the
complex conjugate amplitude (symmetric diagrams) or they may have the
gluon emitted off the quark (antiquark) in the amplitude and off the
antiquark (quark) in the complex conjugate amplitude (asymmetric
diagrams). The gluon production contribution in the diagrams of the
first kind is exactly the same as in pA \cite{kmu} (for a more
detailed review see \cite{yuri1,kw}). The contribution of the second
(asymmetric) type of diagrams differs from the first in: (i) the
interference terms, where the gluon is emitted before the collision in
the amplitude and after the collision in the complex conjugate
amplitude and vice versa; (ii) the late emission time terms, where the
gluon is emitted after the interaction both in the amplitude and in
the complex conjugate amplitude. To understand (i) let us discuss a
specific case of a gluon emitted off the quark line before the
collision in the amplitude and absorbed by the antiquark line after
the collision in the complex conjugate amplitude. There the
interactions with the target nucleons in which at least one of the
exchanged Coulomb gluon lines connects to the quark line cancel by
real-virtual cancellation with the diagrams where that gluon line is
on the other side of the cut. Only the interactions with the gluon and
the antiquark lines survive and could be easily resummed along the
lines outlined above and in \cite{kmu}. The late emission time
diagrams (ii) in the case of asymmetric gluon emission are different
from their symmetric counterparts because the real-virtual
cancellation of interactions here does not happen. The reason is the
color factors which are now different for the ``real'' and ``virtual''
diagrams. Similar to \cite{kmu} one has to consider the last nucleon
interacting with the $q\bar q$ pair before the gluon emission. With
the help of Fierz identity the interaction diagrams for this last
nucleon could be resummed and prove to be diagonal in the color space,
which simplifies their exponentiation. At the end the one gluon
inclusive production cross section for DIS could be written as
\be\label{disxi1}
\frac{d \sigma^{\gamma^* A \rightarrow q{\bar q}GA}_{incl}}{d^2 k \, dy} \, = \, 
\frac{1}{2 \pi^2} \, \int \, d^2 r \, d z \, \Phi^{\gamma^* \rightarrow q\bar q} 
({\underline r}, z) \, \frac{d {\hat \sigma}^{q{\bar q}A}_{incl}}{d^2
k \, dy}({\underline r})
\ee
with
\ben
\frac{d {\hat \sigma}^{q{\bar q}A}_{incl}}{d^2 k \, dy}({\underline x} 
- {\underline y}) \, = \,  
\frac{\as C_F}{\pi^2} \, 
\frac{1}{(2 \pi)^2} \, \int \, d^2 b \, d^2 z_1 \, d^2 z_2 \, 
e^{- i {\underline k} \cdot ({\underline z}_1 - {\underline z}_2)} \,
\left[ \frac{{\underline z}_1- {\underline x}}{|{\underline z}_1- 
{\underline x}|^2} \cdot \frac{{\underline z}_2- {\underline
x}}{|{\underline z}_2- {\underline x}|^2} \right. \, \left( 1 - e^{- ({\underline
z}_1- {\underline x})^2 Q_s^2 /4} \right.
\een
\ben
\left. - e^{- ({\underline z}_2-
{\underline x})^2 Q_s^2 /4} + e^{- ({\underline z}_1- {\underline
z}_2)^2 Q_s^2 /4} \right) + \frac{{\underline z}_1- {\underline
y}}{|{\underline z}_1- {\underline y}|^2} \cdot \frac{{\underline
z}_2- {\underline y}}{|{\underline z}_2- {\underline y}|^2} \, \left(
1 - e^{- ({\underline z}_1 - {\underline y})^2 Q_s^2 /4} - e^{-
({\underline z}_2- {\underline y})^2 Q_s^2 /4} \right.
\een
\ben
+ \left. e^{- ({\underline
z}_1- {\underline z}_2)^2 Q_s^2 /4}\right)
- \frac{{\underline z}_1- {\underline x}}{|{\underline z}_1- {\underline
x}|^2} \cdot \frac{{\underline z}_2- {\underline y}}{|{\underline
z}_2- {\underline y}|^2} \, \left( e^{- ({\underline x}- {\underline y})^2 Q_s^2 /4} - 
e^{- ({\underline z}_1- {\underline y})^2 Q_s^2 /4} - e^{-
({\underline z}_2- {\underline x})^2 Q_s^2 /4} + e^{- ({\underline
z}_1- {\underline z}_2)^2 Q_s^2 /4} \right)
\een
\be\label{disxi2}
\left. - \frac{{\underline z}_1-
{\underline y}}{|{\underline z}_1- {\underline y}|^2} \cdot
\frac{{\underline z}_2- {\underline x}}{|{\underline z}_2- {\underline
x}|^2} \, \left( e^{- ({\underline x}- {\underline y})^2 Q_s^2 /4} -
e^{- ({\underline z}_1- {\underline x})^2 Q_s^2 /4} - e^{-
({\underline z}_2- {\underline y})^2 Q_s^2 /4} + e^{- ({\underline
z}_1- {\underline z}_2)^2 Q_s^2 /4} \right)\right].
\ee
As one can easily check the cross section of \eq{disxi2} goes to zero
in the limit of zero dipole size ${\underline x} \rightarrow
{\underline y}$.

Eqs. (\ref{disxi1}) and (\ref{disxi2}) provide us with the one gluon
inclusive production cross section in DIS calculated in the
quasi-classical approximation and could be used to describe the pion
or minijet production data at HERA.

\section*{Acknowledgements}

I would like to thank Alex Kovner, Larry McLerran and Urs Wiedemann
for many informative and stimulating discussions.  I am grateful to
Stephane Munier for his help in finding a mistake in the earlier
version of the paper. This work was supported in part by the
U.S. Department of Energy under Grant No. DE-FG03-97ER41014 and by the
BSF grant $\#$ 9800276 with Israeli Science Foundation, founded by the
Israeli Academy of Science and Humanities.

\end{document}